\documentclass{ifacconf}

\usepackage{graphicx}      
\usepackage{natbib}        
\usepackage{subfigure}
\usepackage{amsmath} 
\usepackage{amssymb}  
\usepackage{widetext}
\usepackage{algorithm}
\usepackage{algorithmicx}
\usepackage{algpseudocode}
\usepackage[usenames,dvipsnames]{xcolor}

\theoremstyle{break}

\newtheorem{remark}{Remark}

\newcommand{\qi}{{q^{-1}}}

\newcommand{\tha}{ {\hat{\theta}} }

\newcommand{\bm}{\left[\begin{array} }
\newcommand{\mb}{\end{array}\right] }
\newcommand{\bk}{\left\{\begin{array} }
\newcommand{\kb}{\end{array}\right\} }

\newcommand{\algcomment}[1]{{\color{blue}\Comment{{\scriptsize #1}}}}

\algnewcommand{\Initialize}[1]{%
  \State \textbf{Initialize:}
  \Statex \hspace*{\algorithmicindent}\parbox[t]{.8\linewidth}{\raggedright #1}
}

\begin{document}
\begin{frontmatter}

\title{Frequency Separation based Adaptive Feedforward Control for Rejecting Wideband Vibration 
	with Application to Hard Disk Drives\thanksref{footnoteinfo}} 

\thanks[footnoteinfo]{Financial support for this study was provided by a grant from the Advanced Storage Technology Consortium (ASTC).}

\author[First,Second]{Jinwen Pan} 
\author[Second]{Zhi Chen} 
\author[First]{Yong Wang}
\author[Second]{Roberto Horowitz}

\address[First]{University of Science and Technology of China, Hefei, Anhui, 230027, China  
(e-mail: yongwang@ustc.edu.cn)}
\address[Second]{University of California, Berkeley, Berkeley, CA 94706 USA 
(e-mail: horowitz@berkeley.edu)}

\begin{abstract}                

In this paper, a frequency separation based adaptive feedforward control algorithm is developed with the ability to identify the plant and do compensation region by region. In this algorithm, the accelerometer signal is filtered by a series of uniformly distributed bandpass filters to generate a bunch of subband signals which are mutually exclusive in spectrum. In each subband, the corresponding subband signal acts as the feedforward signal and only the frequency response of system in that region needs to be identified, thus a pretty low order model can be expected to have efficient compensation. Starting from the first region, the feedforward control parameters are learned simultaneously with the low order plant model in the same region and then moves to the next region until all the regions are performed. 
\end{abstract}


\end{frontmatter}

\section{Introduction}
The problem of rejecting unknown vibration in a dynamical system is a fundamental control problem, thereby control methods for vibration suppression have been of great interest to researchers both in controls and signal processing communities ever since 1930's [\cite{widrow1985adaptive,pan2016internal}]. Due to the knowledge of the vibration, the controller can be designed either in an feedback fashion or an feedforward fashion. A pictorial example is given in [\cite{nie2009design,pan2016triple}], where non-neglectable NRRO (Non-repeatable runout) exists in a hard disk drive. Usually the NRRO cannot be measured and what is available to engineers is its spectrum, therefore, only feedback controller can be applied to suppress the vibration within its bandwidth. However, if we somehow know the vibration, a feedforward controller can be designed to reduce the vibration without degrade the primary closed loop performance. In [\cite{pan2016dsp}], frequencies of the vibration are available, therefore, both direct [\cite{shahsavari2016adaptive}] and indirect [\cite{shahsavari2014adaptive}] adaptive method are developed based on the known frequencies, known as narrow-band vibration rejection. No matter whether the frequencies are available or not, a feedback control scheme should be designed that adaptively enhances the servo performance at these frequencies, while maintaining the baseline servo loop shape. A novel adaptive multiple narrow-band disturbance observer with a minimum parameterfor selective disturbance cancellation was proposed in [\cite{chen2012minimum}]. For sinusoidal disturbance estimation and compensation, readers may refer to [\cite{bodson1997adaptive,bodson2001active,landau2005adaptive}] and the references therein. 

If we can measure the vibration using a sensor, feedforward vibration rejection is possible no matter the vibration is narrow- or broad-band [\cite{shahsavari2016an}]. In this case, the filtered-x least mean square (Fx-LMS) is known as an effective method to reduce the disturbance effect at the performance side  when the system dynamics are known [\cite{burgess1981active,shahsavari2014adaptive}]. When the system dynamics are unknown, a secondary path modeling is required to perform vibration suppression [\cite{akhtar2007active}], which means the parameters modeling the secondary path also need to be adapted online. There is no computational problem when the vibration is of narrow band and a small number of parameters is sufficient to model the system accurately. However, when the vibration (for instance, in the hard disk drives) has a wide frequency range varying from low frequency to even Nyquist frequency, and the system dynamics are also unknown, a high-order model is required to estimate the real system accurately, which always means computational intensive. When the vibration is periodic with known frequencies, a direct adaptive control method based on frequency separation was developed to attenuate the vibration [\cite{shahsavari2016an}], that can identify the system and learn the feedforward control law region by region. When the vibration is unknown but can be measured using an accelerometer, with the same computational problem, it is impossible to identify the system for the whole frequency range at one time.

With the similar idea as [\cite{shahsavari2016an}], in this paper, we aim to develop a spectrum partition based adaptive feedforward control algorithm that can identify the system and do compensation region by region. The spectral dynamic range is significantly reduced in subband signal than that of fullband signal, thus fast convergence rate can be expected when least mean square (LMS) is performed. The number of parameters adapted in each region is much less than that in methods identifying the fullband system frequency response for full-spectrum compensation. The proposed algorithm can be applied to unknown minimum phase system as well as non-minimum phase system. The adaptive controller is applicable to multi-input single-output (MISO) systems with unknown dynamics. Neither collecting nor processing batches of data is needed. The effectiveness of the algorithm is demonstrated by Matlab simulations with system and noise modeled from realistic hard disk drives.

\section{Problem Statement}
The adaptive controller proposed in this paper is aimed to be implemented in a feedforward fashion, meaning that it is used to augment an existing robustly stable closed-loop system in order to reject special disturbances that can be measured indirectly by an additional sensor -- e.g. by deploying an accelerometer in the case of vibration suppression. These disturbances are not well rejected by the existing baseline controller. 

\begin{figure}[htbp]
	\centering	\includegraphics[width=0.9\linewidth]{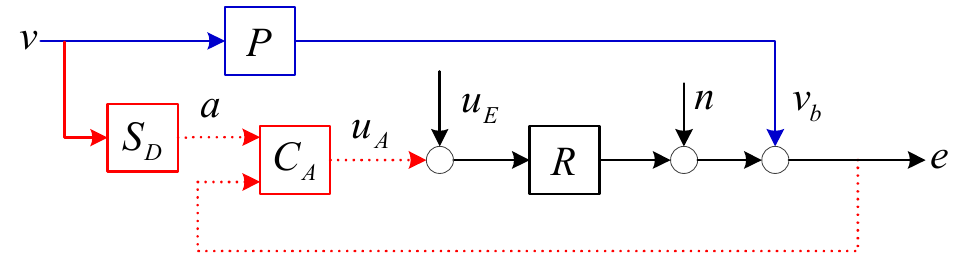}
	\caption{Structure of a stable plant ($R$) augmented by an add--on feedforward controller ($C_A$) responsible for the output disturbance ($ v_b $).}	\label{fig:main_block_diagram_general}
\end{figure}

In this architecture, the baseline feedback controller can be designed  without consideration of this special control task. Moreover, the feedforward controller does not alter the performance of the original control system. To clarify this notion, we use a common Single-Input Single-Output (SISO) plant-controller interconnection shown in Fig.~\ref{fig:main_block_diagram_general} as a running example. 

The block $ R $ in the figure denotes a stable closed-loop system that is formed by a linear time-invariant (LTI) plant and an LTI feedback compensator. The feedforward adaptive  controller, denoted by $C_A$, provides compensation for the special disturbance denoted by ${v_b}(k)$ where $ k $ is the time step index. The source of this disturbance, $v(k)$ contaminates the system at the output after passing through an unknown transfer function (primary path) $P$. The signal $ a(k) $ is an implicit measurement of $v(k)$ by using an additional sensor denoted by ${S_D}$ in the figure. $ u_A $ is the output of the feedforward controller and $ u_E $ is proper designed external excitation signal. One of the main contributions of the controller that will be presented shortly is that it does not require the closed-loop system dynamics as an a-priori information. Since our design does not depend on whether the closed-loop system dynamics are continuous- or discrete-time, we assume that $ R $  is a discrete time systems to make notations simpler and $ z $ is the discrete-time frequency domain variable. 

The special disturbance that should be compensated by the adaptive controller is denoted by ${{v}_b}$, and without loss of generality, we assume that it contaminates the plant output.

An important point to make here is that our plug-in controller design is not limited to this particular interconnection. In general, it does not require any details about the individual components of the closed-loop system and their interconnections. Rather, our design is based on an abstract LTI system from the adaptive control ($u_A$) injection point to the \emph{error} signal ($e$) which is denoted by $R(z)$ as shown in Fig.\ref{fig:main_block_diagram_general}. The only requirement for $ R(z) $ is that it should be stable. It should be emphasized that $ R(z) $ can be non-minimum phase.

When $ v $ is a narrow-band disturbance, the disturbance $ v_b $ is of the same frequency since every component here is LTI. When $ v $ is wideband, a wideband online model for $ R $ is required. When the frequency range is wide, more parameters are needed to describe the system accurately, indicating that more parameters need to be identified adaptively which is too time-consuming to be implemented. In this paper, a frequency separation based algorithm is proposed to solve this problem and make it implementable in real time embedded system. 

\section{Frequency Separation}
\begin{figure}[htbp]
	\centering	\includegraphics[width=1.0\linewidth]{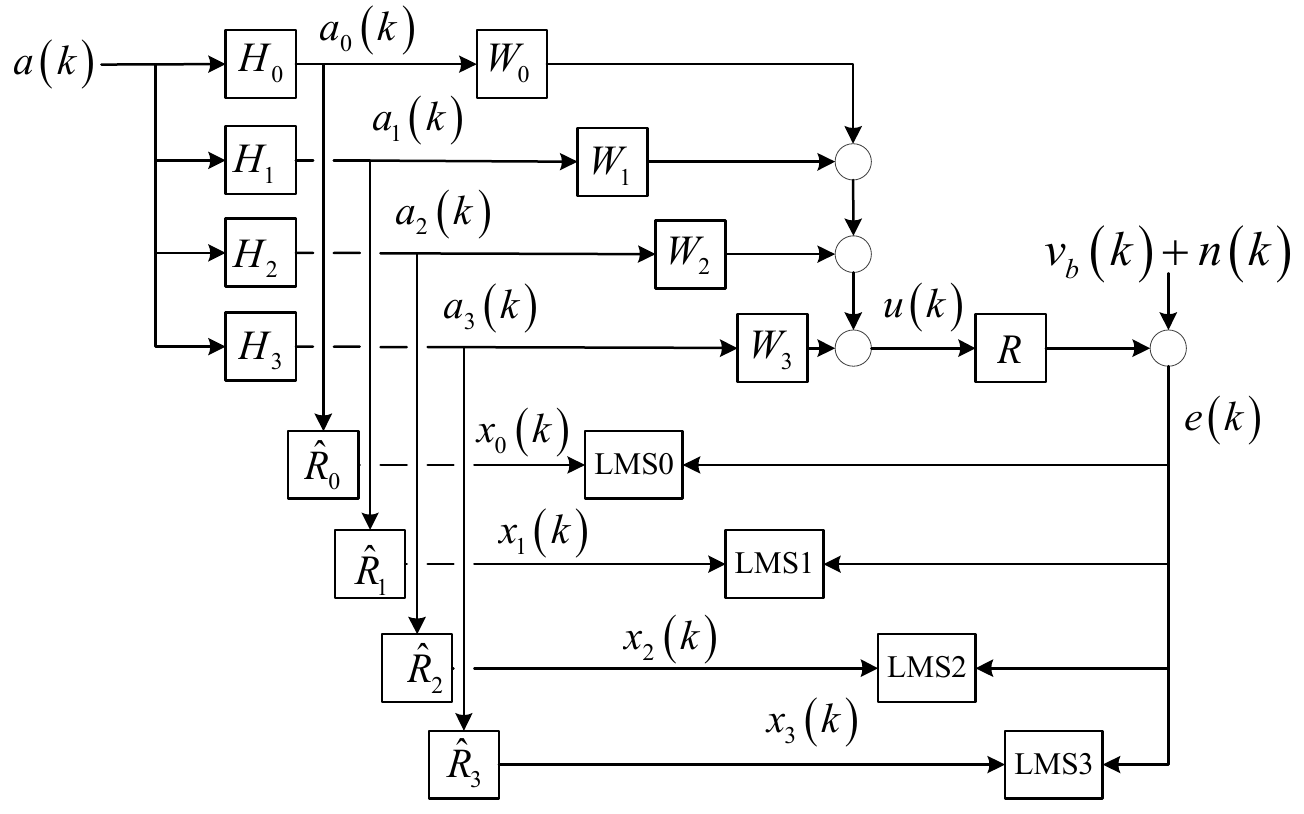}
	\caption{Frequency separation based adaptive feedforward control architecture for a stable plant ($R$) augmented by add--on feedforward controllers ($W_i$) responsible for the output disturbance ($ v_b $) (Four Regions).}
	\label{fig:freq_sepa_block_diagram}
\end{figure}

The frequency separation based adaptive feedforward control architecture for four regions is shown in Fig.~\ref{fig:freq_sepa_block_diagram}, where $ W_i $ is the $ i $-th region FIR feedforward controller, $ H_i $ is the FIR bandpass filter, $ \hat{R}_i(z) $ is the model of the real system $ R(z) $ in $ i $-th region and $ u(k) $ is the injection input to the abstract plant $ R $.

The selection of the regions is critical and is highly related to the application. Generally speaking, the subband should focus on the frequency regions that the vibration located in. The number of regions can be reduced if the vibration is concentrated in a small frequency region. Without loss of generality, here we use four uniformly distributed regions as a running example.

\subsection{SYSTEM IDENTIFICATION}
\begin{figure}[htbp]
	\centering	\includegraphics[width=0.5\linewidth]{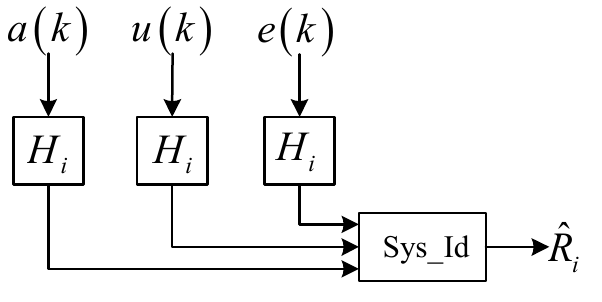}
	\caption{System identification diagram for $ i $-th region.}
	\label{fig:sys_id}
\end{figure}

Since the vibration spectrum is wide, we need the full-band model in order to cancel the vibration effectively. However, full-band model requires mounts of parameters being updated online simultaneously, which is not practicable. As shown in Fig.~\ref{fig:freq_sepa_block_diagram}, here we need to identify the model of the system in each region. Fig.~\ref{fig:sys_id} gives the scheme of how to obtain system model $ \hat{R}_i(z) $. Firstly, we pass all the inputs $ a(k) $, $ u(k) $ and output $ e(k) $ into the same high stop band filter $ H_i $ to obtain the concentrated signals $ a_i(k) $, $ u_i(k) $ and $ e_i(k) $. By synthesizing the concentrated signals, $ R_i(z) $ is obtained by any parameter adaptation algorithm like recursive least squares (RLS). Here we give the procedure of system identification using equations. 

From Fig.~\ref{fig:main_block_diagram_general}, the error $ e(k) $ which we want to minimize is expressed as
\begin{align}\label{eq:error}
	e(k) = R(z)u(k) + v_b(k) + n(k).
\end{align}
where
\begin{align*}
	u(k) = u_A(k) + u_E(k).
\end{align*}

and the vibration $ v_b(k) $ is
\begin{align}
v_b(k) = \frac{P(z)}{S_D(z)}{a(k)}.
\end{align}
For $ i $-th region, multiplying $ H_i(z) $ on both side of (\ref{eq:error}) yields
\begin{align}
H_i(z)e(k) = R(z)H_i(z)u(k) + H_i(z)\left[ v_b(k) + n(k) \right],
\end{align}
which can be simplified as
\begin{align}\label{eq:suberror}
e_i(k) = R(z)u_i(k) + \frac{P(z)}{S_D(z)}{a_i}(k) + \xi_i(k),
\end{align}
where $ \xi_i(k) = H_i(z)n(k)$ and the subband signals $ a_i(k) $, $ u_i(k) $ and $ e_i(k) $ 
\begin{align}\label{eq:bps}
\begin{split}
& a_i(k) = H_i(z)a(k),\\
& u_i(k) = H_i(z)u(k),\\
& e_i(k) = H_i(z)e(k).
\end{split}
\end{align}
Ideally these signals contain only the frequency contents in $ i $-th region if $  H_i(z)$ has infinite stop band. In reality, if $ H_i(z) $ has finite high stop band, $ a_i(k) $, $ u_i(k) $ and $ e_i(k) $ contains mainly the frequency contents in $ i $-th region. Therefore, when these signals are synthesized to identify the system, the model gives the system behavior only in the $ i $-th region. This property will be verified in the numerical examples.

From  (\ref{eq:suberror}), replace $ R(z) $ with $ \frac{B(z)}{A(z)} $, in time domain we have
\begin{align}
\begin{split}
{e_i}\left( k \right) & = {A^*}\left( {{q^{ - 1}}} \right){e_i}\left( k \right) + B\left( {{q^{ - 1}}} \right){u_i}\left( k \right) \\
& + M\left( {{q^{ - 1}}} \right){a_i}\left( k \right) + m\left( k \right),
\end{split}
\end{align}
where $ A^*(\qi) = 1 - A(\qi) $, $ M(\qi) $ is the causal part of $ {P(\qi)}{S^{-1}_D(\qi)} $ and 
\begin{align*}
m(k) = A(\qi){\xi_i(k)} + \left[ \frac{P(\qi)}{S_D(\qi)} - M(\qi) \right] a_i(k).
\end{align*}
With the $ k-1 $ step estimates $ \tha_{Ai}(k-1) $, $ \tha_{Bi}(k-1) $ and $ \tha_{Mi}(k-1) $, the a-priori estimate of the error signal $ e(k)$ is
\begin{align}\label{eq:error:ap}
\begin{split}
\hat{e}^{\circ}_i(k) & = \tha^T_{Ai}(k-1) \phi_{ei}(k) + \tha^T_{Bi}(k-1) \phi_{ui}(k) \\
& + \tha^T_{Mi}(k-1) \phi_{ai}(k),
\end{split}
\end{align}
thereby the a-priori estimation error
\begin{align}
\epsilon^\circ_i(k) = e_i(k) - \hat{e}^\circ_i(k).
\end{align}
Since the $ i $-th frequency region is narrow, a low-order model is sufficient enough to describe the system behavior, which implies that a low-order RLS algorithm is possible to be embedded to identify the parameters $ \tha_{Ai}(k) $, $ \tha_{Bi}(k) $ and $ \tha_{Mi}(k) $. (\ref{eq:sys:ap})- (\ref{eq:sys:fs}) gives the RLS algorithm for $ i $-th region system identification.
\begin{align}
	&\epsilon_i(k) = \frac{\lambda_{Si} {\epsilon^\circ_i(k)}}{\lambda_{Si} + \phi^T_{Si}(k) F_{Si}(k-1) \phi_{Si}(k)}, \label{eq:sys:ap}\\
	&{\tha_{Si}(k)} = {\tha_{Si}(k-1)} + \lambda^{-1}_{Si} F_{Si}(k-1){\phi_{Si}(k)}\epsilon_i(k),\label{eq:sys:tha}
\end{align}
\begin{widetext}
\begin{align}\label{eq:sys:fs}
F_{Si}(k) = \lambda^{-1}_{Si}\left[F_{Si}(k-1) - \frac{F_{Si}(k-1)\phi_{Si}(k)\phi^T_{Si}(k)F_{Si}(k-1)}{\lambda_{Si} + \phi^T_{Si}(k)F_{Si}(k-1)\phi_{Si}(k)}\right],
\end{align}
\end{widetext}
where $ \lambda_{Si} $ is the forgetting factor and
\begin{align*}
& \phi^T_{ei}\left( k \right) = \left[ {e_i\left( {k - 1} \right),e_i\left( {k - 2} \right), \cdots ,e_i\left( {k - {n_{Ai}}} \right)} \right],\\
& \phi^T_{ui}\left( k \right) = \left[ {u_i\left( {k-1} \right),u_i\left( {k - 2} \right), \cdots ,u_i\left( {k - {n_{Bi}}} \right)} \right],\\
& \phi^T_{ai}\left( k \right) = \left[ {a_i\left( {k} \right),a_i\left( {k - 1} \right), \cdots, a_i\left( {k - {n_{Qi}} + 1} \right)} \right],\\
& \phi^T_{Si}(k) = \left[ \phi^T_{ei}(k), \phi^T_{ui}(k), \phi^T_{ai}(k)\right],\\
& \tha^T_{Si}(k) = \left[ \tha^T_{Ai}(k), \tha^T_{Bi}(k), \tha^T_{Mi}(k)\right].
\end{align*}

\subsection{CONTROLLER SYNTHESIS}
The controlller sysnthesis aims to find the implementable form for the control input $ u_{Qi}(k) $ and  finally $ u_A(k) $. Since $ u_{Qi}(k) $ is constructed based on the subband signal $ a_i(k) $, they are almost uncorrelated with one another, thus the total control input can be computed using
\begin{align*}
u_A(k) = \sum \limits_{i=1}^{N} { u_{Qi}(k) }.
\end{align*}

For the $ i $-th region, when the plant $ R(z) $ is known, we can compute the filtered signal $ x_i(k) $ as
\begin{align*}
x_i(k) = R(z)a_i(k),
\end{align*}
and the controller parameters for the $ i $-th region are updated as
\begin{align}\label{eq:ctrl:tha}
&{\tha_{Qi}(k)} = {\tha_{Qi}(k-1)} + \lambda^{-1}_{Qi} F_{Qi}(k-1){\phi_{Qi}(k)}e(k),
\end{align}
with $ F_{Qi}(k) $ updated as
\begin{widetext}
\begin{align}\label{eq:ctrl:fq}
F_{Qi}(k) = \lambda^{-1}_{Qi}\left[F_{Qi}(k-1) - \frac{F_{Qi}(k-1)\phi_{Qi}(k)\phi^T_{Qi}(k)F_{Qi}(k-1)}{\lambda_{Qi} + \phi^T_{Qi}(k)F_{Qi}(k-1)\phi_{Qi}(k)}\right],
\end{align}
\end{widetext}
where $ \lambda_{Qi} $ is the forgetting factor, $ \tha_{Qi}(k) $ is the estimated 
parameter vector of $ W_i(z) $ shown in Fig.~\ref{fig:freq_sepa_block_diagram} and 
\begin{align*}
& \phi^T_{Qi}\left( k \right) = \left[ {x_i\left( {k} \right),x_i\left( {k - 1} \right), \cdots ,x_i\left( {k - {n_{Qi}}} \right)} \right].
\end{align*}
However, $ R(z) $ is not available when the system dynamics is unknown. Fortunately, in the previous step, we have the estimated low order system model $ \hat{R}_i(z) $, then the filtered signal $ x_i(k) $ can be alternatively computed as 
\begin{align}\label{eq:xi}
x_i(k) = \hat{R}_i(z) a_i(k),
\end{align}
which is always implementable. The feedforward control input for the $ i $-th region is
\begin{align}\label{eq:uQi}
u_{Qi}(k) = \tha^T_{Qi}(k-1) \phi_{ai}(k).
\end{align}

\subsection{PARAMETER ADAPTATION PROCEDURE}
For parameter adaptation, one way is to update all the parameters of $ \hat{R}_i(z) $ and $ W_i(z) $ in different region simultaneously. However, it is almost impossible to do so because of too much computation in one control period. An alternative way that takes benefit of frequency separation is to update the parameters of $ R_i(z) $ and $ W_i(z) $ region by region. Specifically, we identify $ \hat{R}_0(z) $ and $ W_0(z) $. After their parameters converging, we froze the parameters and move to $ \hat{R}_1(z) $ and $ W_1(z) $ and so on until all parameters in all regions converge. The whole algorithm is described in \textbf{Algorithm.~\ref{alg:direct_rls_SecondaryPath}}.

\begin{algorithm}[htbp]
\caption{FREQ\_SEPA}\label{alg:direct_rls_SecondaryPath}
	\begin{algorithmic}[1]
		\Procedure{FREQ\_SEPA}{$ a(k) $, $ e(k) $, $ N $}
        \State $ N $: number of regions
        \While {$ i < N $} 
        \State Read $a(k)$ and update $a_i(k)$									\algcomment{\eqref{eq:bps}}	
        \State Compute $u_{Qj}(k)$, $ j = 0,1,\ldots, i $ and $ u_A(k) = u_{Q0}(k)+ \cdots + u_{Qi}(k) $ and inject $ u(k) $ to the system 							\algcomment{\eqref{eq:uQi}}
        \State Read $ e(k) $ and update $ e_i(k) $ \algcomment{\eqref{eq:bps}}	
        \State Identify $ \hat{R}_i(z) $  \algcomment{\eqref{eq:error:ap} - \eqref{eq:sys:fs}}
        \State Compute $ x_i(k) $ and update $ \phi_{Qi}(k) $ \algcomment{\eqref{eq:xi}}
        \State Update $ \tha_{Qi}(k) $ until it converges \algcomment{\eqref{eq:ctrl:tha} - \eqref{eq:ctrl:fq}}
        \If {$ \tha_{Qi}(k) $ converges} 
        \State Fix $ \hat{R}_i(z) $ and $ W_i(z) $
        \State $ i = i + 1 $ 
        \EndIf
        \EndWhile
		\EndProcedure
	\end{algorithmic}
\end{algorithm}

\begin{remark}
\textbf{Algorithm.~\ref{alg:direct_rls_SecondaryPath}} is designed specifically for the configuration shown in Fig.~\ref{fig:main_block_diagram_general} where the abstract system $ R(z) $ is single-input-single-output (SISO). However, the \textbf{Algorithm.~\ref{alg:direct_rls_SecondaryPath}} also works for MISO system because of its frequency separation property when the actuators are performed based on their frequency characteristics. Examples will be given in the numerical examples. 
\end{remark}

\section{Numerical Examples}
\begin{figure}[htbp]
	\centering	
    \includegraphics[width=1.0\linewidth]{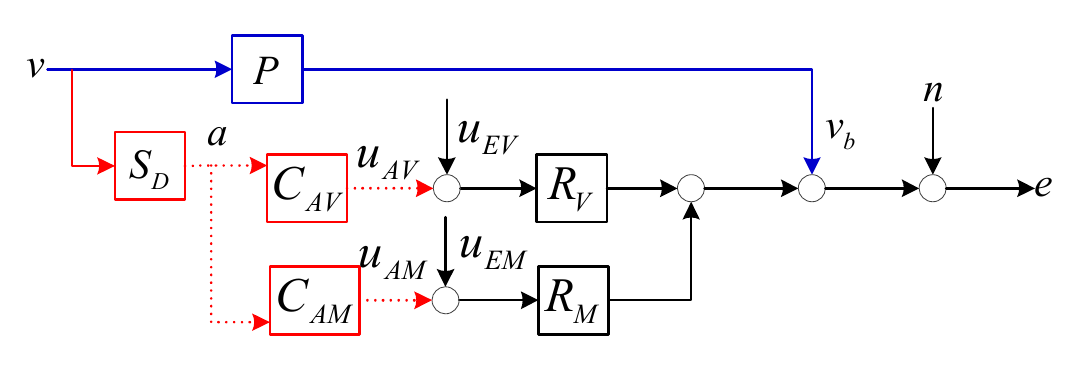}
	\caption{Dual-stage feedforward vibration rejection configuration.}
	\label{fig:dual_stage_ffwd}
\end{figure}

The effectiveness of the proposed algorithm is verified by designing a vibration suppression controller for a dual-stage hard disk drive servo system. Most of current HDDs are equipped with accelerometer that can measure vibration in one or multiple directions [\cite{pannu1998adaptive}]. The block diagram in Fig.~\ref{fig:main_block_diagram_general} can be adopted to represent a dual-stage HDD servo controller in track-following mode as shown in Fig.~\ref{fig:dual_stage_ffwd}. The blocks $R_V$ and $R_M$ refer to the Voice Coil Motor (VCM) and Micro Actuator (MA) respectively. $ C_{AV} $ and $ C_{AM} $ represent the adaptive feedforward controller for VCM and MA respectively. $ u_{EV} $ and $ u_{EM} $ are possible external excitation signals for system identification designed for VCM and MA. The signal $n$ corresponds to airflow disturbance, runout, and measurement noise lumped together. The output signal $e(k)$ denotes the measured \emph{position error signal} (PES). The main objective is to design adaptive controllers that receive the accelerometer signal in accordance with the PES, and feedforward proper control signals to the inputs of VCM and MA respectively to suppress vibration $v_b(k)$.

Adaptive algorithms are plausible methods for HDDs since an exact dynamics of the actuator is not known for each individual unit. Moreover, temperature variation and deterioration can increase uncertainty over time [\cite{pan2016triple}]. In this case study, we do not use any information about the system dynamics or the feedback controller, and assume that the only available signals are the position error $e(k)$ and the acceleration measurement $a(k)$.

\begin{figure}[htbp]
	\centering	
    \includegraphics[width=1.0\linewidth]{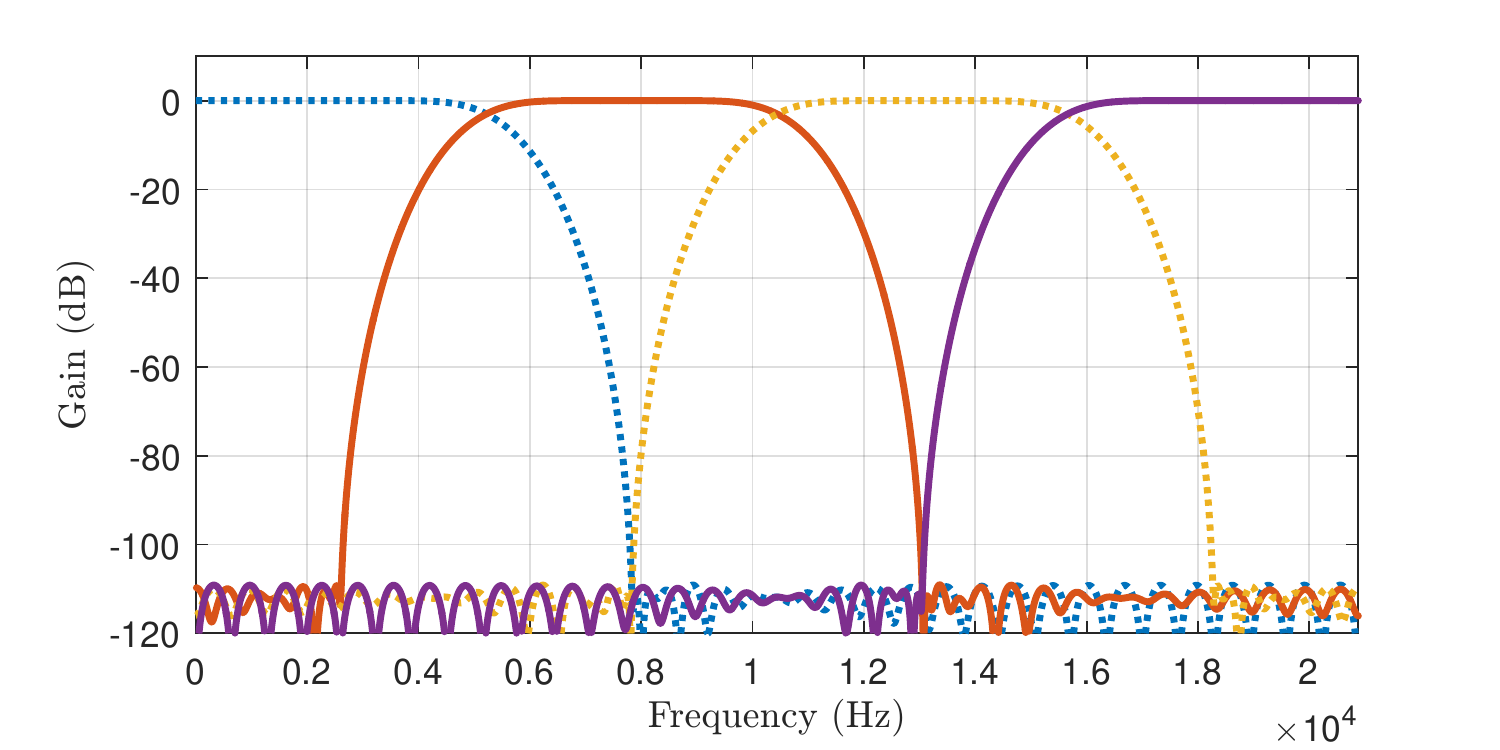}
	\caption{Uniformly distributed bandpass filters using DCT (Four Regions, $ 64 $-th order FIR filter).}
	\label{fig:dct}
\end{figure}

In this simulation, four uniform regions from low frequency to Nyquist frequency (here the sampling frequency is $ F_s = 41, 760\text{Hz} $, therefore the Nyquist frequency is $ \frac{1}{2}{F_s} = 20, 880\text{Hz} $) are considered for vibration rejection, two regions for VCM and two regions for MA. Fig.~\ref{fig:dct} gives the related uniformly distributed bandpass filters. These bandpass filters are of high stop band and generated as analysis filter banks using cosine modulation [\cite{lee2010subband}]. Here in our simulation, all bandpass filters are designed as $ 64 $-th order FIR filter with an attenuation around $ 110\text{dB} $. Then $ a_i(k) $, $ i = 0,1,2,3 $ are in almost mutually exclusive spectral bands. They are almost frequency independent and uncorrelated with one another. 

\begin{figure}[htbp]
	\centering
	\includegraphics[width=0.9\linewidth]{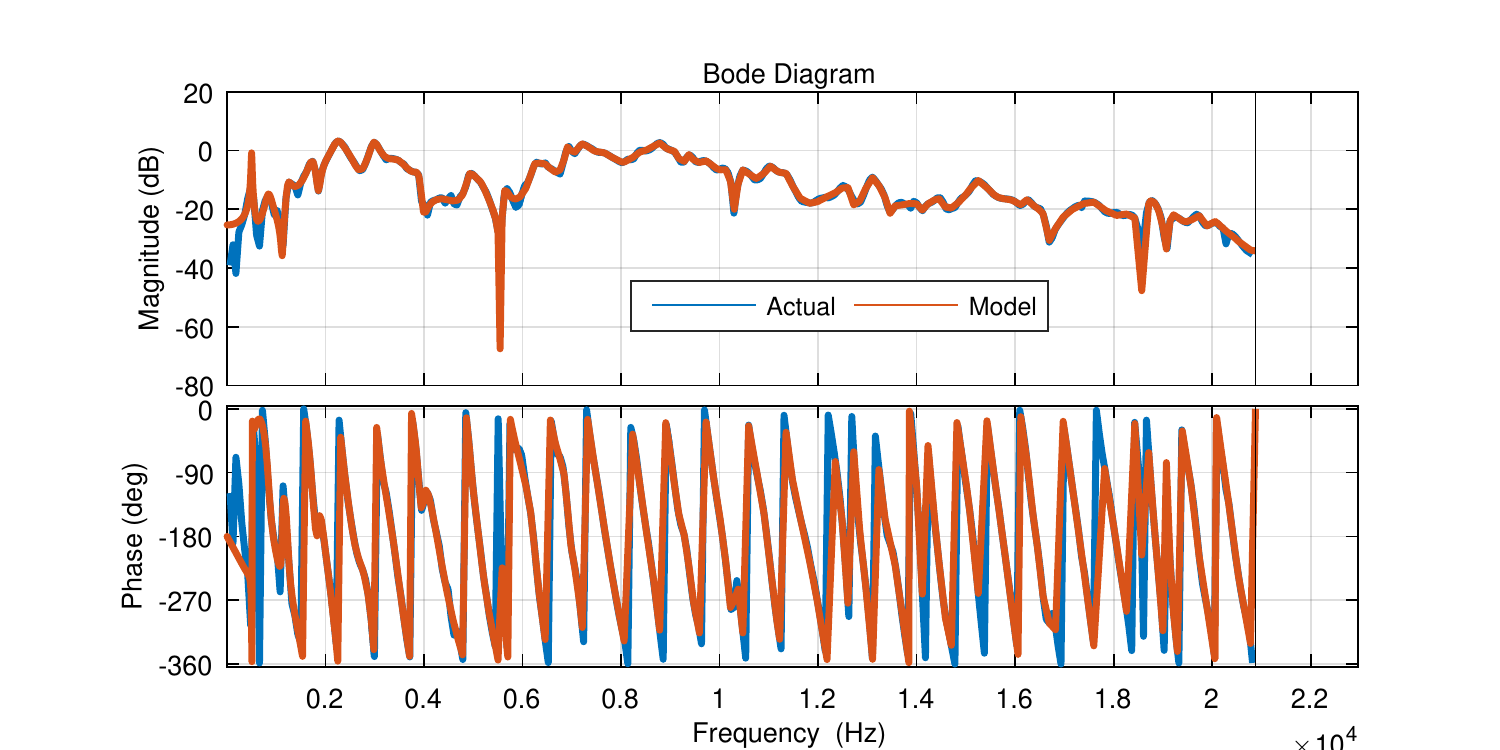}
	\caption{Frequency response comparison of the identified model and the actual sensor ($S_D(z)$). }
	\label{fig:bode_shaker_acc}
\end{figure}

\begin{figure}[htbp]
	\centering
	\includegraphics[width=0.9\linewidth]{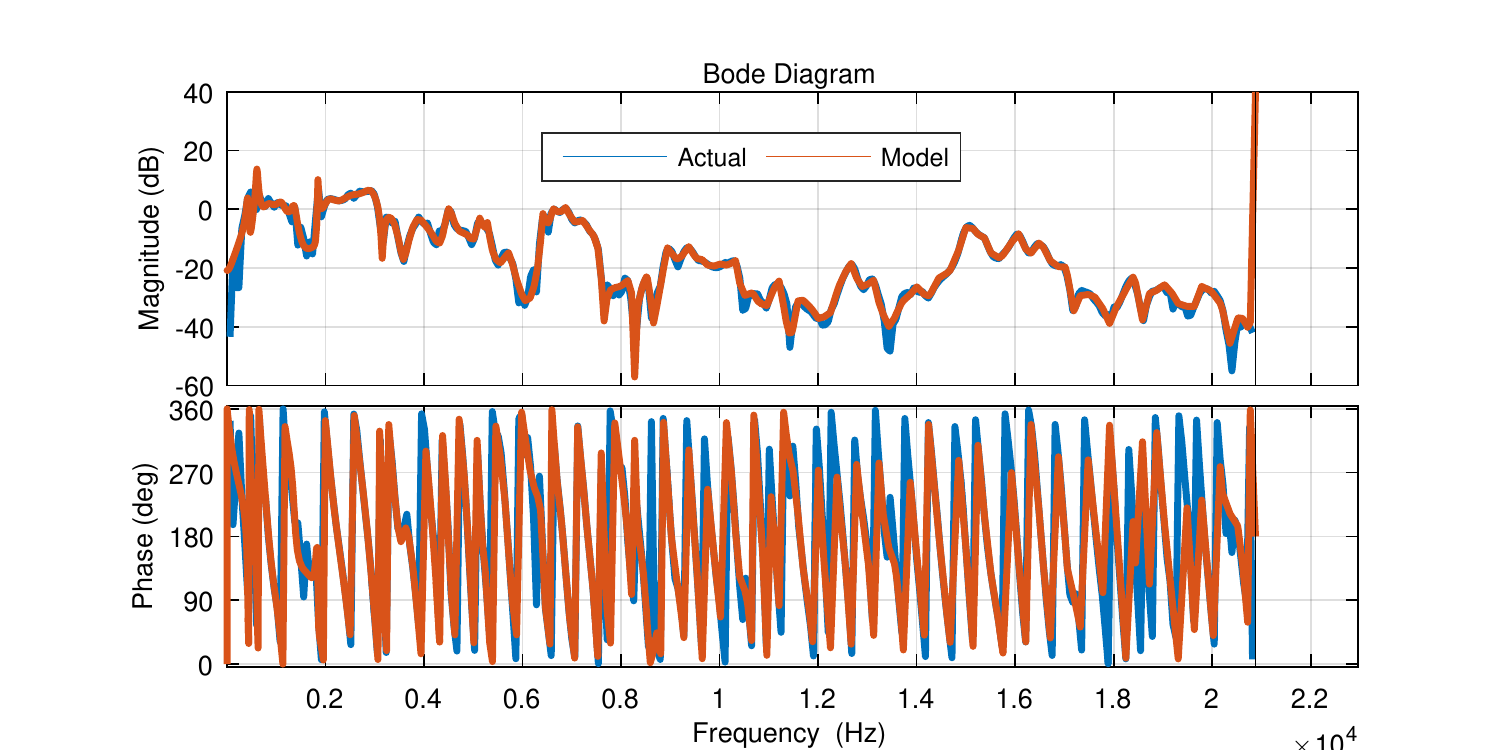}
	\caption{Frequency response comparison of the identified model and the actual primary path ($P(z)$).}
	\label{fig:bode_shaker_pes}
\end{figure}

In our simulation, the primary path $ P(z) $, sensor path $ S_D(z) $, VCM model $ R_V(z) $ and MA model $ R_M(z) $ are all modeled from real experimental frequency response data, where the models $ S_D(z) $ and $ P(z) $ with their frequency response data are given in Fig.~\ref{fig:bode_shaker_acc} and Fig.~\ref{fig:bode_shaker_pes}, respectively. All the system and controller orders used in the simulation are listed in Table.~\ref{tb:order}. Notably, here the system orders for each $ R_i(z) $ are not necessarily same with each other. The only requirement is that it should be low-order and at the same time, can describe the system behavior precisely. RRO and NRRO are also modeled from realistic measurements. In this way, our simulation will be very close to the experiment results and can be used as reference. 

\begin{table}[htbp]
\begin{center}
\caption{System and Controller Orders}\label{tb:order}
\begin{tabular}{cccccc}
\hline
System & Order & System & Order & System & Order \\\hline
$ R_V(z) $ &  50 (IIR)  & $ R_0(z) $ &  5 (IIR) & $ R_1(z) $ &  5 (IIR)  \\
$ R_M(z) $ &  17 (IIR)  & $ R_2(z) $ & 5 (IIR) & $ R_3(z) $ & 5 (IIR)\\
$ W_0(z) $ & 3 (FIR) & $ W_1(z) $ & 3 (FIR)& $ W_2(z) $ & 3 (FIR)\\
$ W_3(z) $ & 3 (FIR) & $ P(z) $ & 125 (IIR) & $ S_D(z) $ & 100 (IIR) \\ \hline
\end{tabular}
\end{center}
\end{table}

\begin{figure}[htbp]
\centering
\subfigure{\includegraphics[width=0.48\linewidth]{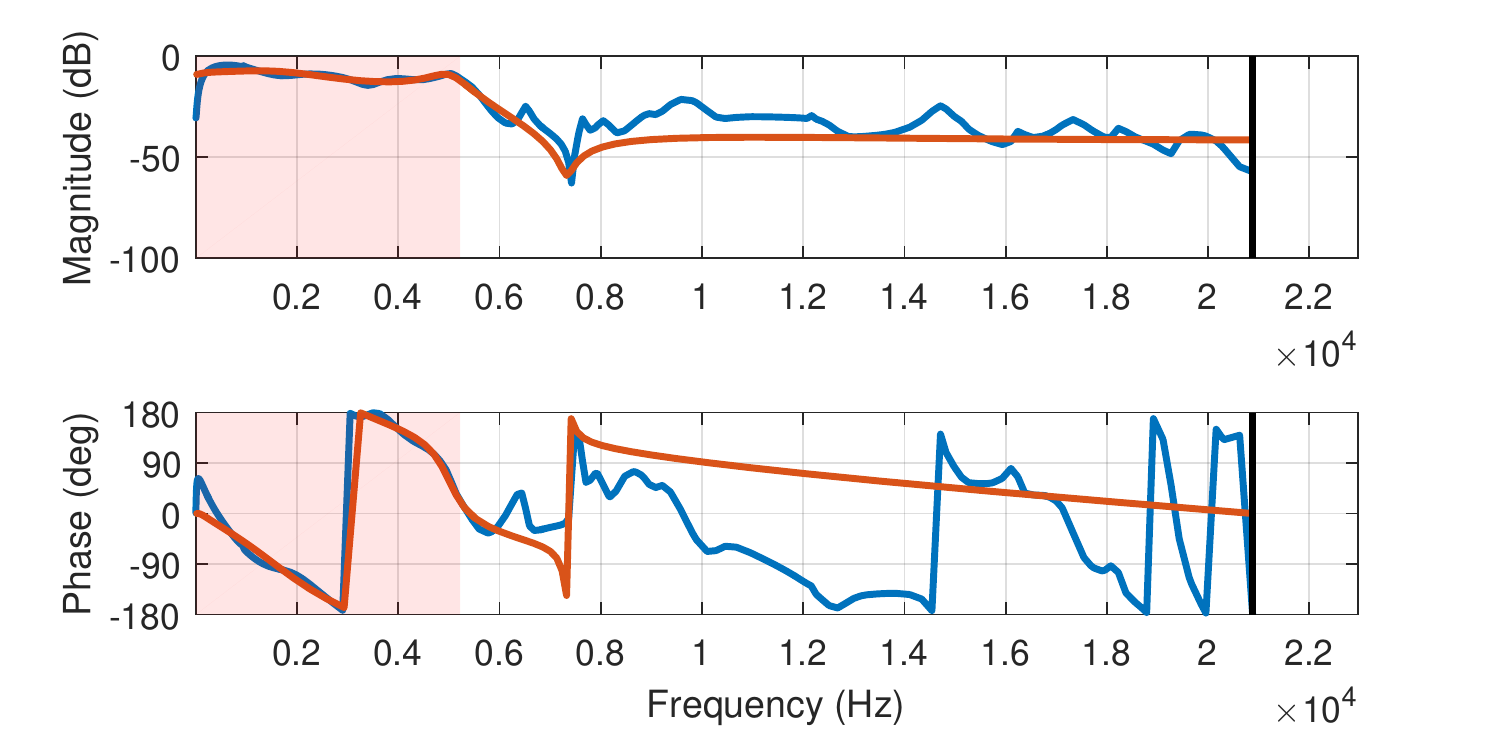}}
\subfigure{\includegraphics[width=0.48\linewidth]{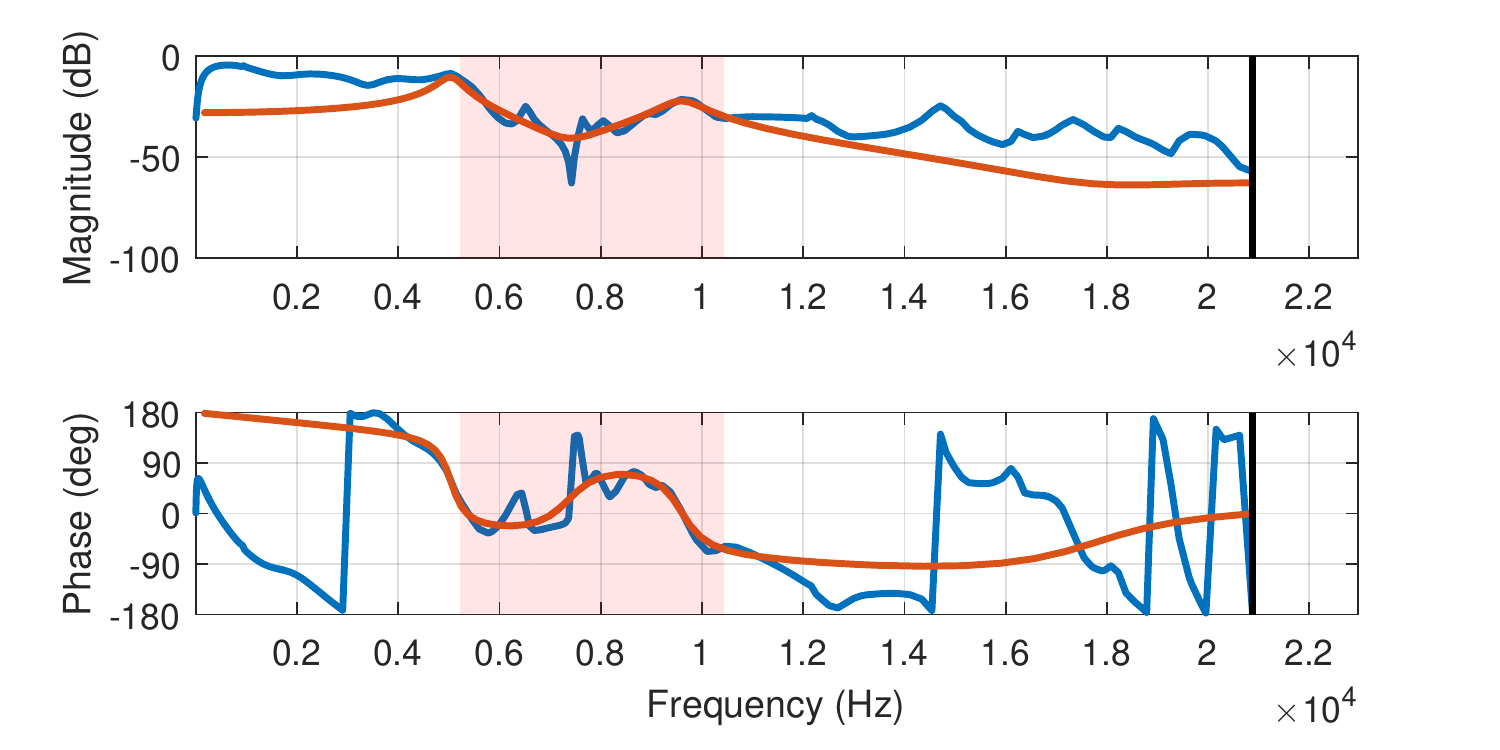}}
\subfigure{\includegraphics[width=0.48\linewidth]{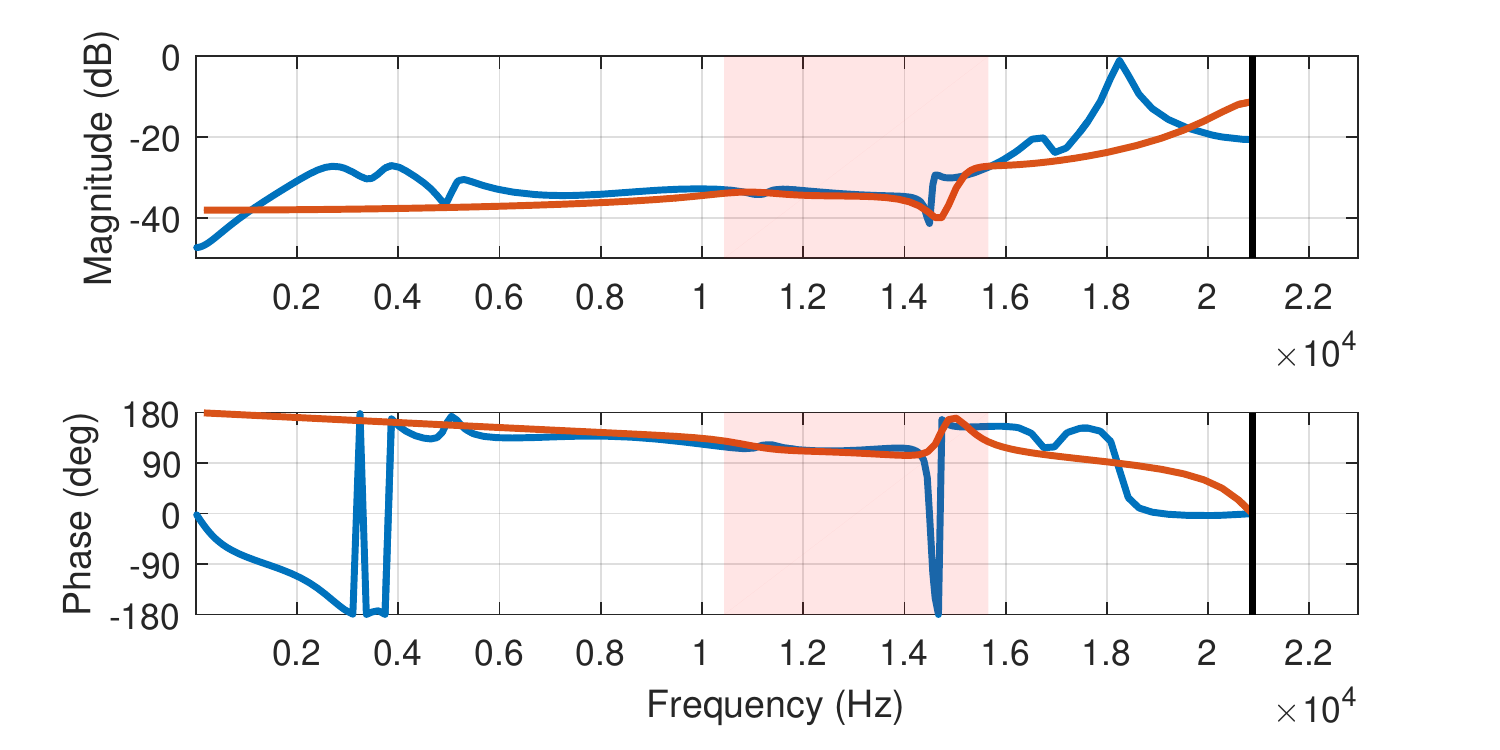}}
\subfigure{\includegraphics[width=0.48\linewidth]{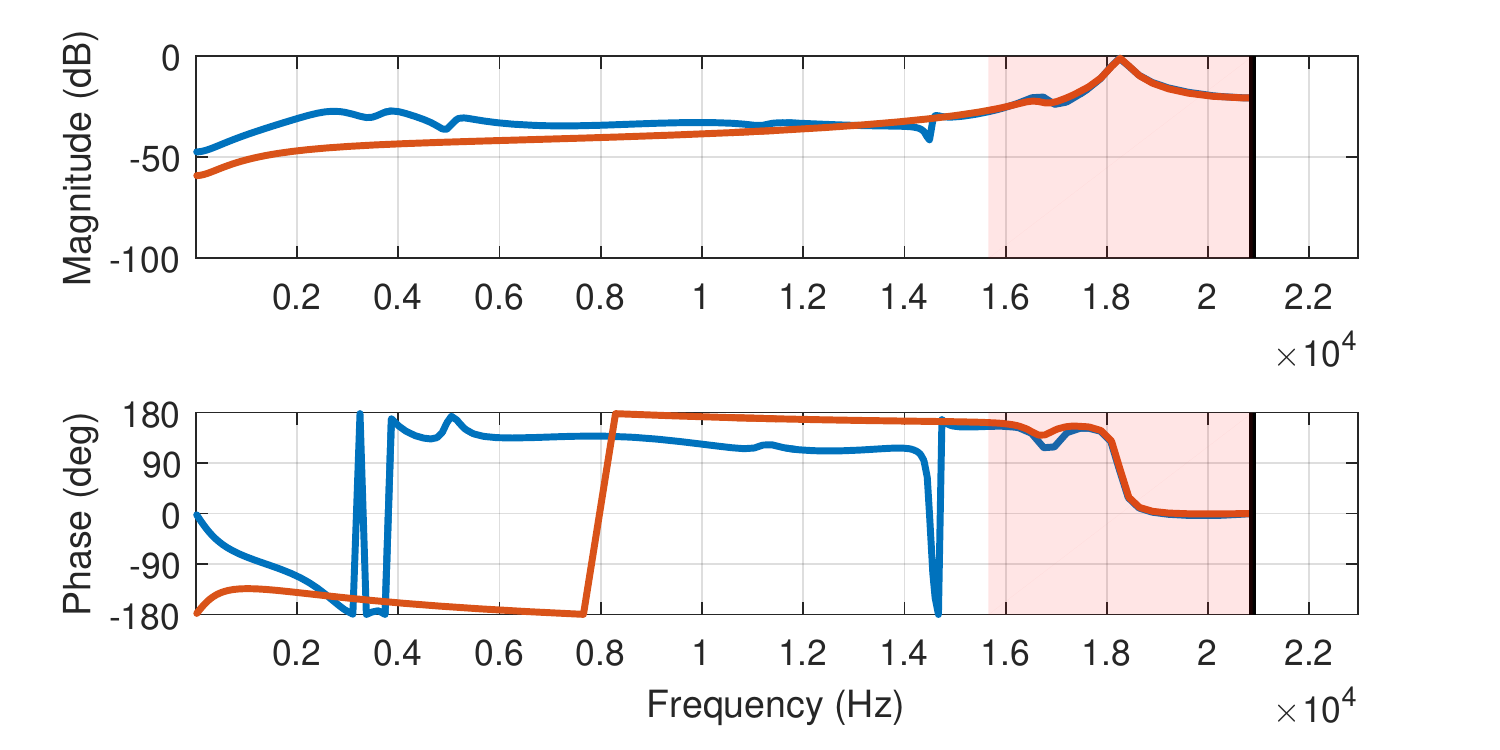}}
\caption{Frequency response comparison between estimates and real plants: Upper Left (Blue: real VCM, Red: estimate of region $ 0 $); Upper Right (Blue: real VCM, Red: estimate of region $ 1 $); Lower Left (Blue: real MA, Red: estimate of region $ 2 $); Lower Right (Blue: real MA, Red: estimate of region $ 3 $).}
\label{fig:tABhbode}
\end{figure}

\begin{figure}[htbp]
	\centering	\subfigure{\includegraphics[width=0.48\linewidth]{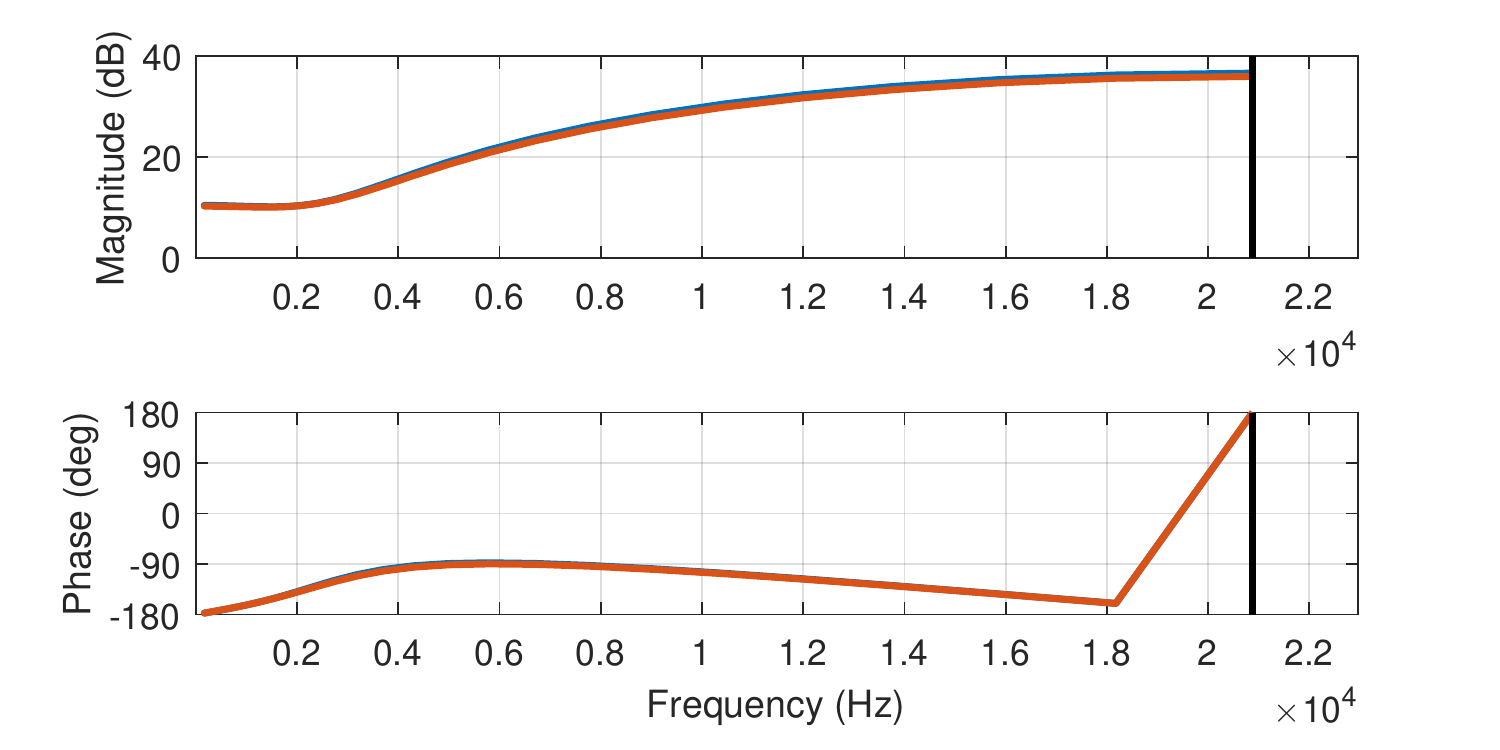}}	\subfigure{\includegraphics[width=0.48\linewidth]{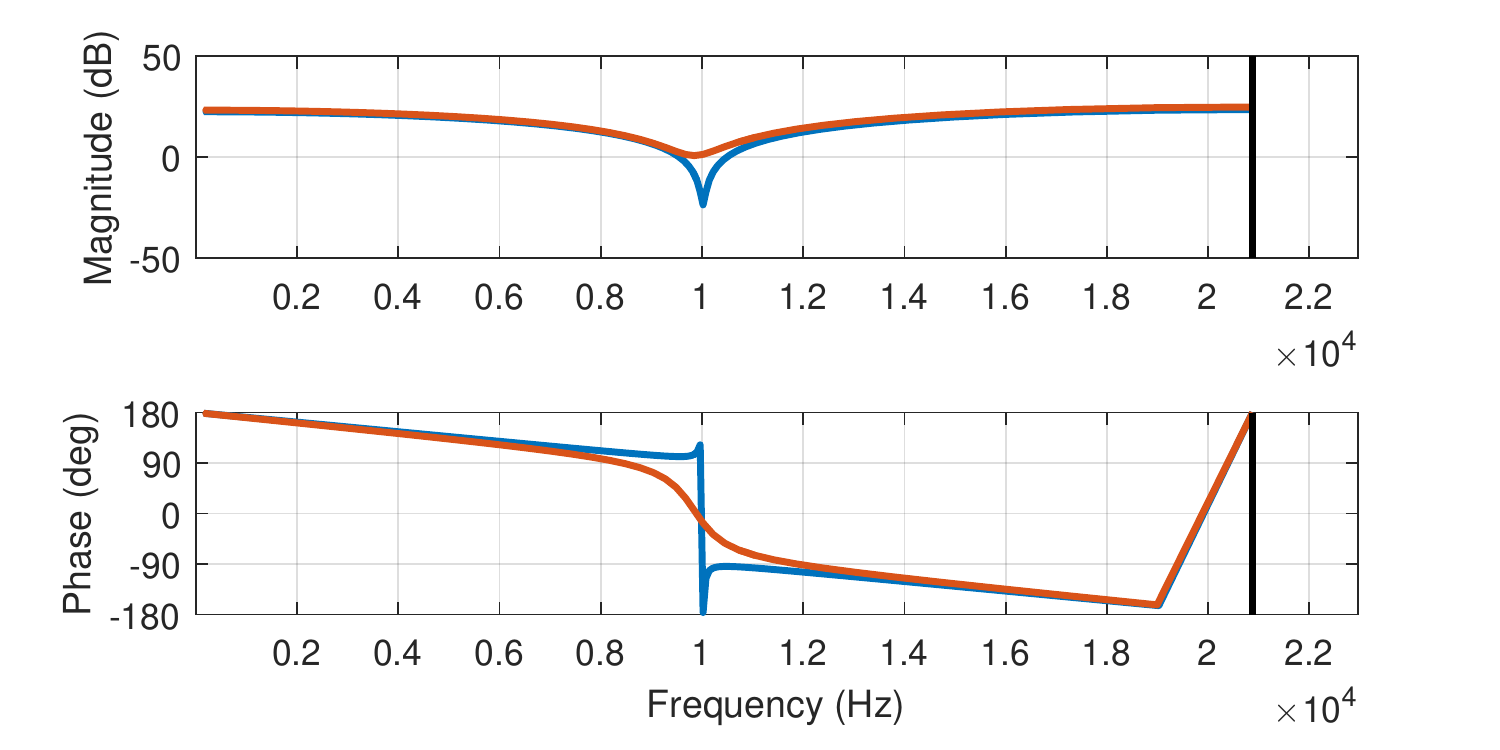}}	\subfigure{\includegraphics[width=0.48\linewidth]{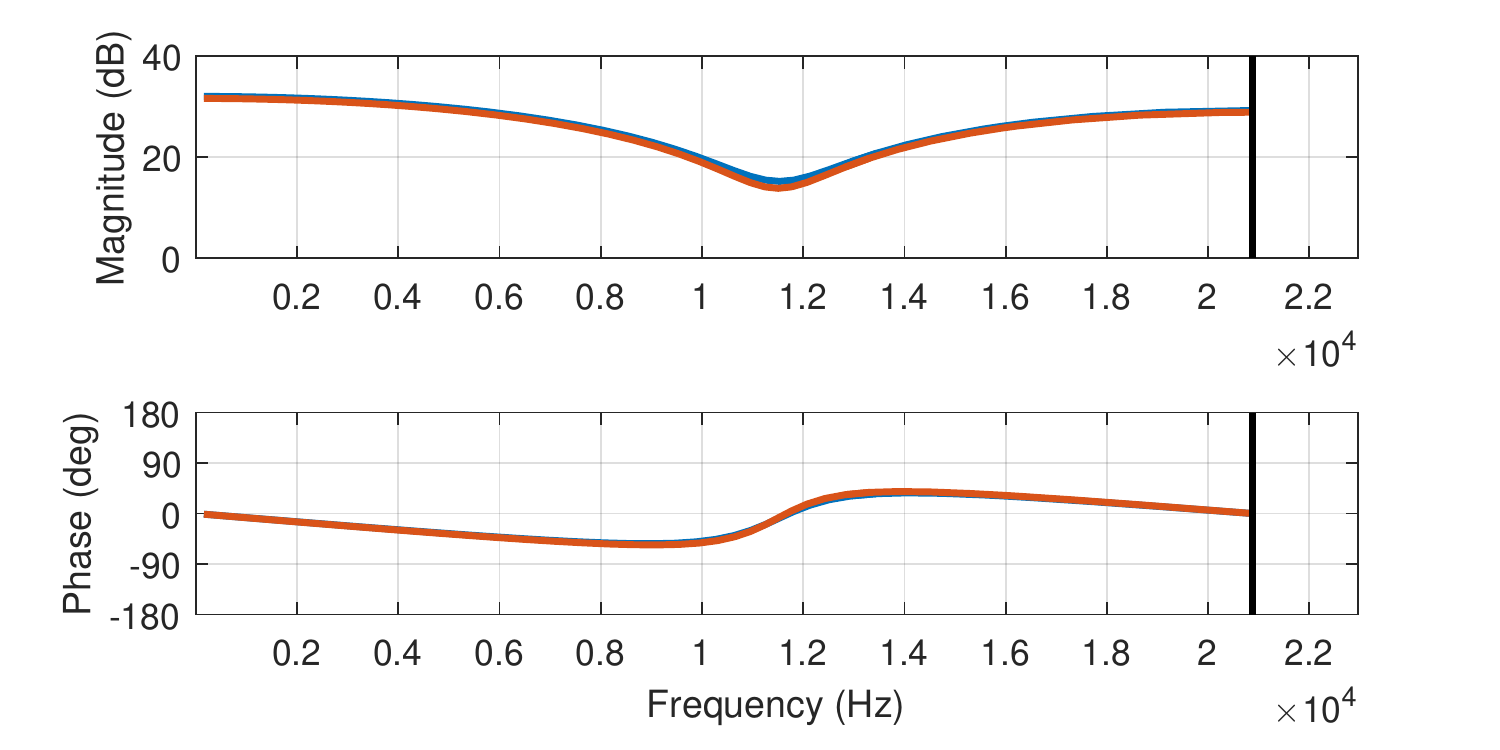}}	\subfigure{\includegraphics[width=0.48\linewidth]{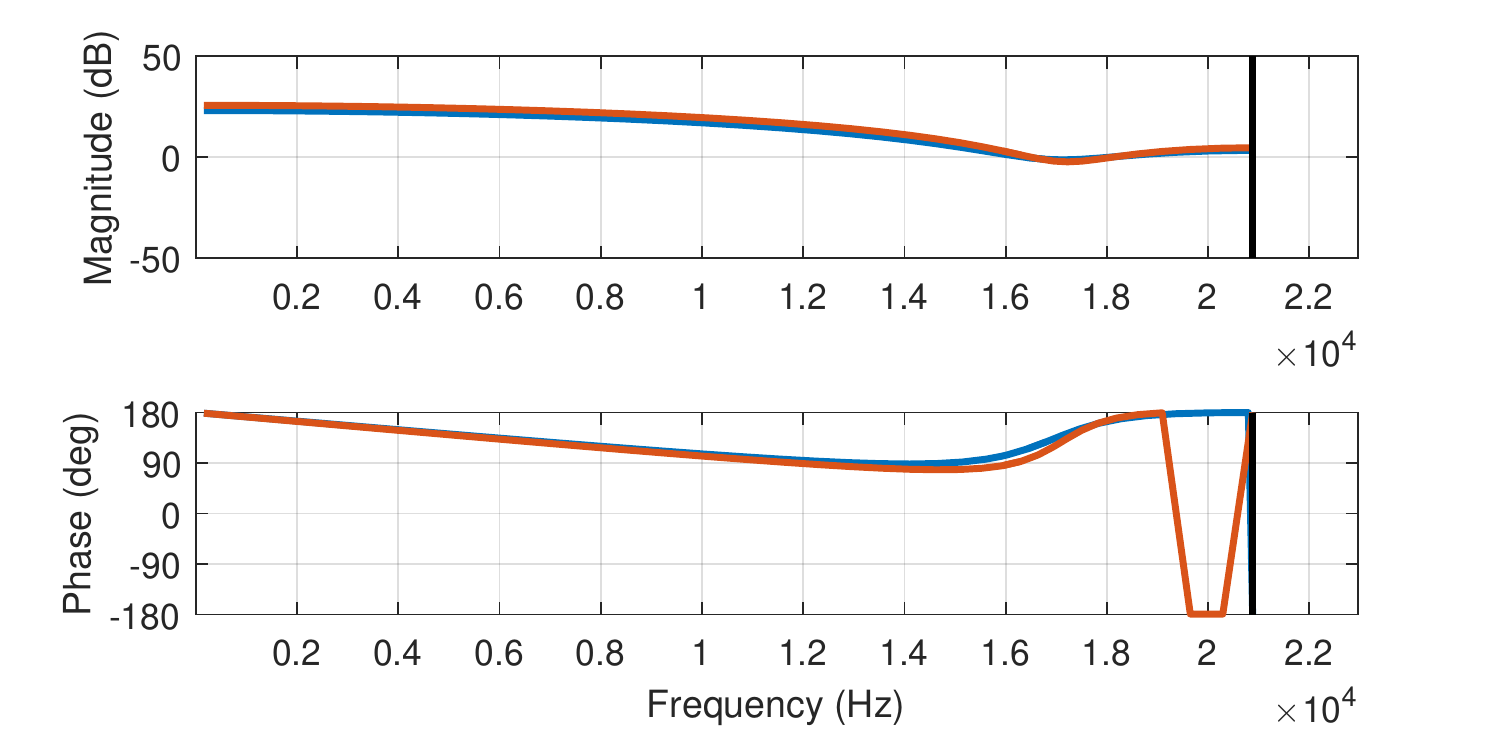}}
	\caption{Frequency response comparison between estimates and optimal controller: Upper Left (Blue: optimal , Red: estimate of $ C_{AV} $ in region $ 0 $); Upper Right (Blue: optimal, Red: estimate of of $ C_{AV} $ in region $ 1 $); Lower Left (Blue: optimal, Red: estimate of of $ C_{AM} $ in region $ 2 $); Lower Right (Blue: optimal, Red: estimate of of $ C_{AM} $ in region $ 3 $).}
	\label{fig:tQhbode}
\end{figure}

The simulation results are shown in Fig.~\ref{fig:tABhbode} and Fig.~\ref{fig:tQhbode}. System identification for four regions are given in Fig.~\ref{fig:tABhbode}, where the blue ones are actual model and red ones are the estimated model. The shadow in each sub figure is the frequency range that fully covering that specific region. It can be observed that in the considered region, the system is modeled very close to actual system using a low order IIR model, which verifies the above mentioned property. Fig.~\ref{fig:tQhbode} gives the frequency response of ideal $ W_i(z) $ and their estimations. It can be seen that all feedforward controllers converge very close to their optimal value.

\section{Conclusions}
A spectrum partition based feedforward algorithm is proposed to solve the parameters burden problem caused by wide vibration frequency range. The number of parameters adapted in each region is much less than that of methods identifying the system frequency response, making it possible for whole spectral vibration compensation. The proposed algorithm is applicable to unknown minimum phase systems as well as non-minimum phase systems. The adaptive controller is applicable to multi-input single-output (MISO) systems with unknown dynamics. No collecting or processing batches of data is needed. The effectiveness of the algorithm is demonstrated by comprehensive Matlab simulations with system as well as noise modeled from a realistic dual-stage hard disk drive which is a non-minimum phase MISO system.

\begin{ack}
Financial support for this study was provided by a grant from the Advanced Storage Technology Consortium (ASTC).
\end{ack}

\bibliography{ifacconf}
                                                   







\end{document}